\documentclass[iop,12pt]{emulateapj}
\usepackage{amssymb,amsmath}
\usepackage{natbib} 
\usepackage[dvipsnames]{color}
\def\s{{\rm\,s}}
\def\m{{\rm \,m}}

\def\km{{\rm\,km}}

\def\kpc{{\rm\,kpc}}

\def\pc{{\rm\,pc}}
\def\cm{{\rm\,cm}}

\def\K{{\rm\, K}}
\def\mK{{\rm\,mK}}
\def\refnew#1{\,(\ref{#1})}
\def\arcsec{{\rm \, arcsec}}
\def\max{{\rm max}}
\def\esu{{\rm\, esu}}

\def\gff{{\bar{g}_{\rm ff}}}
\def\GHz{{\, \rm GHz}}
\def\Gpc{\,\rm{Gpc}}
\def\DM{{\rm DM}}
\def\rel{{\rm rel}}
\def\ms{{\rm \, ms}}
\def\DMunit{\,\mathrm{pc\,cm^{-3}}}
\def\dif{{\rm dif}}
\def\HII{{\rm HII}}
\def\th{{\rm th}}
\def\Jy{{~\rm Jy}}

\def\MHz{{\rm MHz}}

\begin{document}

\title{Physical Constraints On Fast Radio Bursts}
\author{Jing Luan \& Peter Goldreich}
\email{jingluan@caltech.edu}
\affiliation{California Institute of Technology, Pasadena, CA 91125, US}

\begin{abstract}
Fast Radio Bursts (FRBs) are isolated, $\ms$ radio pulses with dispersion measure (DM) of order $10^3\DMunit$. Galactic candidates for the DM of high latitude bursts detected at $\GHz$ frequencies are easily dismissed.  DM from bursts emitted in stellar coronas are limited by free-free absorption and those from HII regions are bounded by the nondetection of associated free-free emission at radio wavelengths. Thus, if astronomical, FRBs are probably extra-galactic.  FRB 110220 has a scattering tail of $\sim 5.6\pm 0.1 \ms$.  If the electron density fluctuations arise from a turbulent cascade, the scattering is unlikely to be due to propagation through the diffuse intergalactic plasma.  A more plausible explanation is that this burst sits in the central  region of its host galaxy.  Pulse durations of order $\ms$ constrain the sizes of FRB sources implying high brightness temperatures that indicates coherent emission. Electric fields near FRBs at cosmological distances would be so strong that they could accelerate free electrons from rest to relativistic energies in a single wave period. 
\end{abstract}

\maketitle

\section{Introduction}

FRBs are single, broad-band pulses with flux densities $S_\nu \sim \Jy$ and durations $\Delta t\sim \ms$ detected at $\nu\sim \GHz$ \citep{2007Sci...318..777L, 2013Sci...341...53T}.  They were discovered by de-dispersing data collected by the Parkes multi-beam radio telescope during pulsar searches.  Thus far there are no reports of FRBs detected by other radio telescopes.  The procedure followed in the detection of FRBs is similar to that which led to the discovery of rotating radio transients (RRATs, \cite{2006Natur.439..817M}) now firmly identified as sporadically active pulsars.  \cite{2013Sci...341...53T} report the detection of four high-galactic-latitude ($>40^\circ$) FRBs with DM of several hundred $\DMunit$, well above the contribution expected from our Galaxy \citep{2002astro.ph..7156C}. It has become popular to attribute these large DMs to propagation through the intergalactic plasma indicating source distances $d\sim \Gpc$.  

Currently, it is unclear whether FRBs herald the discovery of a new type of astronomical source or merely that of an unidentified source of noise.  The strongest argument supporting the astronomical origin of FRBs is the precise degree to which arrival times of individual pulses follow the $\nu^{-2}$ law that characterizes the propagation of radio waves through a cold plasma.  Some pulses detected in searches for FRBs are clearly terrestrial although their origin is unknown.  These have been named Perytons.  It is notable that the classification of the Lorimer burst \citep{2007Sci...318..777L} remains controversial, although if it is a FRB it would be the first and brightest  of those detected. For the remainder of this paper, we cast aside our doubts and proceed as through FRBs are bonafide astronomical signals. Interest in detecting additional FRBs with other radio telescopes is high \citep{2013MNRAS.tmpL.147L, 2013ApJ...776L..16T}), so we expect their true nature to be revealed soon.  In \S \ref{extra-galactic}, we show that the DMs of FRBs cannot arise from propagation through a stellar corona or a galactic HII region.  Then in \S \ref{scattering}, we argue that intergalactic electron density fluctuations are unlikely to provide the angular deflections deduced from the temporal scattering tail resolved in FRB 110220. \S \ref{radio-sky} shows that the contribution of FRBs to the brightness of the radio sky is negligible. In \S \ref{brightness-temperature}, we discuss the high brightness temperatures of FRBs and assess the possibility that FRBs are signals  beamed at Earth by advanced civilizations. \S \ref{strongE} discusses the strength of the electric fields of FRBs in terms of their ability to accelerate free electrons to relativistic energies in 
one radio wave period. We summarize our results in \S \ref{discussion} and briefly comment on possible emission mechanisms for FRBs.

\section{Sources are probably extra-galactic}\label{extra-galactic}
In this section, we discuss two galactic candidates to produce $\DM$ for FRBs. One is a stellar corona, suggested by \cite{2014MNRAS.tmpL...2L}; the other is an HII region. We then demonstrate that neither can account for the large $\DM$ of FRBs. Thus, the sources of FRBs are probably extra-galactic.

\subsection{Free-free absorption in stellar coronas}\label{flarestar}
\cite{2014MNRAS.tmpL...2L} proposed that FRBs originate from flares on main-sequence stars and that the DM arises from propagation through the stellar corona. This proposal has the attractive feature of greatly reducing the source luminosity with respect to that required for an unspecified extragalactic source.  Nevertheless, free-free absorption limits a stellar corona's DM to be well below that of FRB's.

In the Rayleigh-Jeans limit, $h\nu\ll k_BT$, the free-free absorption coefficient including stimulated emission reads \citep{1978ppim.book.....S}
\begin{eqnarray}
\alpha&=&\frac{4}{3}\left(\frac{2\pi}{3}\right)^{1/2} \frac{Z^2 e^6n_en_i\gff}{c\ m_e^{3/2}\left(k_BT\right)^{3/2}\nu^2}\, ,\\
\gff&=& \frac{\sqrt{3}}{\pi}\left[\ln\left(\frac{(2k_B T)^{3/2}}{\pi e^2m_e^{1/2}\nu}\right)-\frac{5\gamma}{2}\right]\, ,
\label{eq:alphaff}
\end{eqnarray}
where $\gff$ is the Gaunt factor and $\gamma=0.577$ is Euler's constant. Other symbols are standard: $m_e$ is the electron mass, $e$ is the electron charge, and $n_e$ and $n_i$ are the number densities of electrons and ions.  For cosmic abundances and in the temperature range of interest here, it suffices to evaluate $\alpha$ for a pure hydrogen plasma, i.e., $Z=1$ and $n_e=n_i$.  

For a homogenous medium, the optical depth for free-free absorption is $\tau\sim \alpha s\propto n_e^2 s$, where $s$ is the path length along the line of sight through the medium. Since $\DM=n_e s$, we express $n_e$ in terms of $s$ and $\tau$. Then $\tau\lesssim 1$ sets an upper limit on $\DM$,
\begin{eqnarray}
\DM&\sim & \frac{3^{3/4}(m_e k_B T)^{3/4}(cs)^{1/2}\nu}{2^{5/4}\pi^{1/4}e^3 \gff^{1/2}}\, .\label{DM-upper}
\end{eqnarray}
For $k_B T\lesssim GMm_p/R$ the base of the corona would be in quasi-hydrostatic equilibrium.  Since density drops rapidly with height in an isothermal atmosphere, we replace $s$ in Eq.\refnew{DM-upper} by the scale height $2 k_B T/(m_pg)$, and $k_B T$ by $G Mm_p/R$ to obtain
\begin{eqnarray}
\DM&\sim & \frac{3^{3/4}c^{1/2}m_e^{3/4}(G M m_p)^{3/4}\nu}{2^{3/4}\pi^{1/4}e^3R^{1/4}\gff^{1/2}}\\
&\sim & 50\left(\frac{M}{M_{\sun}}\right)^{3/4}\left(R\over R_\sun\right)^{-1/4}\DMunit
\label{DM}
\end{eqnarray}
which is much smaller than the DMs of FRBs. 

A hotter corona could provide a larger $\DM$.  If free to expand, it would essentially be a stellar wind even close to the photosphere.  For simplicity, the wind is taken to have constant velocity and constant temperature.  These approximations are not entirely consistent because a supersonic isothermal wind would slowly accelerate as its density declined.  This inconsistency leads us to overestimate dispersion measure relative to free-free absorption because the former and latter are proportional to density and density squared. 
At constant radial velocity, $n_e(r)\sim n_e(R)(R/r)^2$.
\begin{equation}
\DM\sim \int_R^{\infty} n_e dr\sim n_e(R) R\, .
\end{equation}
From Eq.\refnew{eq:alphaff},  $\alpha= Cn_e^2/(k_B T)^{3/2}$,  
\begin{equation}
\tau\sim \int_R^{\infty} \alpha dr\sim \int_R^{\infty}  \frac{C n_e^2}{(k_B T)^{3/2}} dr\sim \frac{C n_e(R)^2\ R}{3 (k_B T)^{3/2}}\, .
\end{equation}
The power carried by the wind would be\footnote{In calculating the power needed to drive the wind, we neglect the heat that must be supplied in order to overcome the cooling effect of adiabatic expansion. }
\begin{eqnarray}
P_w&\sim& 4\pi m_p n_e(R) R^2 v_{\th}^3\sim\frac{2^6\pi^{3/2}e^6 \gff  \DM^3 }{3^{5/2}c m_e^{3/2} m_H^{1/2} \nu^2 \tau}\\
&\gtrsim&  40 L_\sun \left(\frac{\DM}{10^3\DMunit}\right)^3 \left(\frac{\nu}{1\GHz}\right)^{-2} ,
\end{eqnarray}
where we have expressed $n_e$ and $s$ in terms of $\DM$ and $\tau$. The $\gtrsim$ on the second line follows from setting $\tau\lesssim1$.  Clearly a coronal wind cannot carry more energy than the luminosity of its star could provide.   Thus even the lowest DM measured for the FRB's reported by \cite{2013Sci...341...53T}, DM$\sim 553\DMunit$, could not  arise from propagation through a coronal wind from the flare stars discussed by \cite{2014MNRAS.tmpL...2L}.  

A hotter corona might be confined by a strong magnetic field provided the magnetic stress is comparable to the gas pressure. Under this condition, the ratio of the electron cyclotron frequency to the plasma frequency would be
\begin{equation}
\frac{\omega_{ce}}{\omega_p}\approx \left(k_B T\over m_ec^2\right)^{1/2}\, .
\label{freqratio}
\end{equation}
Then the dispersion relation for radio waves would depend on $\omega/\omega_{ce}$ in addition to $\omega/\omega_p$ which might cause the frequency dependence of the pulse arrival times to deviate by more than the limits
set by observations of FRBs. 

Numerical results given above are scaled with respect to parameters pertaining to the sun.  Typical flare stars are lower main sequence dwarfs for which $R\propto M^{0.9}$ and
$L\propto M^{3.4}$ \citep{1991Ap&SS.181..313D}.  Application of these relations only strengthens our conclusion that the $\DM$s of FRBs cannot be attributed to passage of radio waves through coronas.

Before moving on, we offer a few additional comments about radio emission from flare stars. This topic has been discussed for more than half a century, starting with
the paper "Radio Emission from Flare Stars" by \cite{1963Natur.198..228L}.  To our knowledge, no bursts sharing the common properties of FRBs have been reported. Moreover, the most frequently studied radio flare stars are close by.  For example, AD Leonis and YZ Canis Minoris are at distance of $\sim 5\pc$ and $\sim 6\pc$, respectively.  These two stars figure prominently, and in most cases exclusively, in each of the papers on radio flares referenced in \cite{2014MNRAS.tmpL...2L} and their strongest bursts barely reach the level of $1\Jy$ that is typical of FRBs.  Dynamic spectra of radio bursts from AD Leonis observed with wide bandwidth and at high time resolution at Arecibo \citep{2006ApJ...637.1016O, 2008ApJ...674.1078O} do not resemble those of FRBs.  Pulses suffering dispersion induced time delays should only show negative frequency drifts. But the histogram of these bursts (Figure 4a in \cite{2006ApJ...637.1016O}) exhibits both positive and negative frequency drifts and is symmetric about zero drift with half width at half maximum of $\sim 3\times 10^{-4}\, s/\MHz$. Note that a $\DM\sim 20\DMunit$ produces a negative frequency drift rate of similar magnitude.

\subsection{HII region}
An HII region is another candidate to account for the DM of a galactic FRB.
A lower limit on $s$ is deducible from Eq.\refnew{DM-upper}. With $T\sim 10^4\K$, $s\ga 0.2\pc$. The angular size of such an HII region at $500\pc$ is $d\theta_{\HII}\sim 80\arcsec$. At $1.4\GHz$, the $64\m$, Parkes telescope's beam size is $d\theta\sim\lambda/D\sim 20\cm/64\m\sim 650\arcsec$. Thus the antenna temperature of such an HII region would be $T_A=T_{\HII}\times(d\theta_{\HII}/d\theta)^2\sim 150\K$. The sensitivity of Parkes at $1.4\GHz$ for a $270\s$ integration time is $0.6\mK$ for $10\sigma$ detection of FRBs (cf. Parkes user guide). Thornton et al. would have recognized an HII region with these properties in the data they search for FRBs.

The bottom line from this section is that neither a stellar corona nor an HII region is a plausible candidate for the high $\DM$s of FRBs'. Thus FRBs are likely to be extragalactic.

\section{Temporal Scattering}\label{scattering}
We follow conventions developed in the investigation of angular and temporal scattering in the interstellar medium \citep{1990ARA&A..28..561R} and adopt the Kolmogorov spectrum,  $\delta n_e/n_e\sim (\ell/L)^{1/3}$, for electron density fluctuations on scale $\ell$ where $\ell_{\min}\leq \ell<L$.  Moreover, we assume that this spectrum is associated with a turbulent cascade in which sonic velocity fluctuations are present at outer scale $L$.\footnote{Conclusions reached in this section depend on the assumption that the electron density fluctuations arise from a turbulent cascade.}  Finally, the scattering is described by projecting the phase differences that 
accumulate along the line of sight between source and observer onto a thin screen located midway between them.   For a source at distance $d$, we obtain
\begin{equation}\label{Dphi}
\Delta\phi \sim \frac{\ n_e e^2 d^{1/2} \ell^{5/6}\lambda}{\pi  m_e c^2 L^{1/3}}\, .
\end{equation}

We are concerned with strong scattering which requires $\Delta\phi>1$.  Then the scattering angle 
\begin{equation}
\Delta\theta \sim \frac{\lambda}{\ell}\Delta\phi\propto \ell^{-1/6}\, 
\label{thscat}
\end{equation}
is dominated by the smallest scale for which $\Delta\phi\gtrsim 1$. For sufficiently small $\ell_{\min}$, this is the diffraction scale at which $\Delta\phi\sim 1$;
\begin{equation}
\ell_{\dif}\sim \left(\frac{\pi m_e c^2}{e^2 n_e \lambda}\right)^{6/5}\frac{L^{2/5}}{d^{3/5}}\, .
\label{elldif}
\end{equation}
Otherwise it is $\ell_{\min}$.  The temporal delay, 
\begin{equation}
\Delta t_{sc}\sim  \frac{d}{c}(\Delta\theta)^2\, 
\label{tsc}
\end{equation}
is expressed as
\begin{equation}
\Delta t_{sc}\sim \left\{\begin{array}{cc}
\frac{d}{c}\left(\frac{\lambda}{\ell_{\dif}}\right)^2\propto \lambda^{4.4}\, ,& \ell_{\min}\leq \ell_{\dif}\, ; \\
& \\
\frac{d}{c}\left(\frac{\Delta\phi \lambda}{\ell_{\min}}\right)^2\propto \lambda^4\, , &\ell_{\min}> \ell_{\dif}\, .
 \end{array}\right.
\label{Deltsc}
\end{equation}

FRB 110220 exhibits a well-resolved exponential tail with $\Delta t_{sc}\sim 5.6\pm 0.1\,\ms$ that has been attributed to plasma scattering \citep{2013Sci...341...53T}. 
Unfortunately, the data is not quite good enough to distinguish between the two cases given in Eq.\refnew{Deltsc} \citep{2013Sci...341...53T}. But both restrict the outer scale
to be less than
\begin{eqnarray}
L_\max&\sim& \left(\frac{e^2\ n_e}{\pi m_e c^2}\right)^3\, \frac{\lambda^{11/2}\ d^{11/4}}{(c\Delta t_{sc})^{5/4}}\sim 10^{13} \left(\frac{d}{\Gpc}\right)^{11/4}\nonumber\\
&&\nonumber \\
&&\times \left(\frac{n_e}{10^{-7}\,\cm^{-3}}\right)^{3}\left(\frac{\Delta t_{sc}}{\ms}\right)^{-5/4}\, \cm\, .
\end{eqnarray}
$L_\max$ is an impossibly small outer scale for extragalactic turbulence.\footnote{In a clumpy IGM with volume filling factor $f$, $L_\max$ would be larger by $f^{-3/2}$.} Sonic velocity perturbations dissipate their bulk kinetic energy into heat on the timescale over which sound waves cross the outer scale.  This would imply a doubling of the IGM temperature over several months since the cooling rate is comparable to the Hubble time.  

Based on the argument given above, it seems unlikely that propagation through the diffuse IGM could make a measurable contribution to the scattering tail of a FRB.  Indeed, an outer scale of order $10^{24}\cm$ is required to reduce the turbulent heating rate to a level compatible with the cooling rate.  With this value, $\Delta t_{sc}\lesssim 10^{-12}\s$ for $d\sim \Gpc$.  Previously, \cite{2013ApJ...776..125M} expressed doubt that propagation through the diffuse IGM could produce discernible scattering tails for FRBs.  However, they failed to recognize the incompatibility of a small $L_\max$ with regulation of the IGM's temperature.

\section{Contribution to Radio Sky}\label{radio-sky}
\cite{2013Sci...341...53T} estimate a FRB event rate of $\sim 10^4\mathrm{~sky}^{-1}\mathrm{day}^{-1}\sim 0.1\s^{-1}$. Given characteristic flux densities of a Jansky and durations of a few milliseconds, FRBs add about $10^{-9}\K$ to the radio background at $1.4\GHz$.\footnote{Unless FRBs are extragalactic, this is merely their contribution to the radio background near our position in the Galaxy.} This value is dwarfed by contributions of $2.7\K$ from the CMB and
even by minor additions from the galactic halo, the galactic plane and extragalactic radio sources.  According to \cite{2013ApJ...776...42S}, these account for $0.79\K$, $0.3\K$ and $0.14\K$ respectively at $1.4\GHz$.

\section{Brightness Temperature}\label{brightness-temperature}
FRBs are not angularly resolved, and thus their brightness temperatures ($T_B$) are unknown. However, the duration of a pulse, $\Delta t$, constrains the linear size of the source and thus its angular size at a fixed distance.  Relativistic beaming is a complication. Radiation emitted from a spherical shell expanding with Lorentz factor $\Gamma$  is beamed into a solid angle $\Delta\Omega\sim \Gamma^{-2}$. Arrival times of photons emitted simultaneously spread by $R/(c\Gamma^2)$ permitting a source size as large
\begin{equation}
R\lesssim c\Delta t\Gamma^2\, .
\label{sourcesize}
\end{equation}
Consequently, the brightness temperature in the observer's frame is
\begin{eqnarray}
T_B&\simeq&  \frac{S_{\nu} d^2}{k_B \Gamma^2\nu^2 \Delta t^2}\\
&\sim& \frac{10^{36}\K}{\Gamma^2}\left(\frac{S_{\nu}}{\Jy}\right)\left(\frac{d}{\Gpc}\right)^2\left(\frac{\nu}{\GHz}\right)^{-2}\left(\frac{\Delta t}{\ms}\right)^{-2}.\nonumber
\end{eqnarray}
Even at $d\sim \kpc$, $T_B\sim 10^{23}/\Gamma^2\K$.  Incoherent broad-band radio emission from strong astronomical sources is usually synchrotron radiation.  Upper limits
on $T_B$ are typically no larger than a few times $10^{13}\K$ \citep{2005AJ....130.2473K}.  This is consistent with an upper limit on $T_B\sim 10^{12}\K$ in the source frame set by the Compton  catastrophe \citep{1992apa..book.....F} with somewhat higher values due to beaming in AGN jets.  

Terrestrial communications at radio wavelengths invariably involve coherent sources.  Could FRBs be signals beamed at us from advanced civilizations?  Might negatively chirped $\ms$ pulses be transmitted to facilitate their detection?  Advanced civilizations would know the power of de-dispersing radio signals to investigate pulsars.  They would also be aware
of planets in their neighborhoods and have identified those with atmospheres suitable for, or perhaps even modified by, biological life. After all, this information will be available to us before the end of this century. 

How might advanced civilizations configure antennas to transmit narrow beams?  Arrays of small telescopes are preferable to large filled apertures and also limit capital costs. With baseline, $b$, and transmitted power, $P_T$, the flux density of a broad-band signal received at distance $d$ would be
\begin{equation}
S_\nu\sim \left(b\over cd\right)^2\nu P_T\, .
\label{eq:Snur}
\end{equation}
Recasting the above equation with $S_\nu$ scaled by $\Jy$ as appropriate for a FRB yields
\begin{equation}
P_T\sim 10^9\left(b\over 10^3\km\right)^{-1}\left(d\over\kpc\right)^2\left(\nu\over \GHz\right)^{-1}\left(S_\nu\over \Jy\right){\rm watt}\, ,
\label{eq:PT}
\end{equation}
a modest power requirement even by current terrestrial standards.\footnote{Scattering by
plasma density fluctuations in the interstellar medium of typical paths would not increase the angular width of
these beams.} 

Accounting for a burst arrival rate at Earth $\sim 0.1\s^{-1}$ is the most challenging part of this exercise.  With only a handful of detected FRBs, the fraction of planets hosting an advanced civilization
might be quite modest.  But then, the Earth must have been recognized as a particular object of interest to target. If this hypothesis has merit, the positions from which bursts arrive should eventually repeat. That would provide a lower limit to the number of our more advanced neighbors.

\section{Strong Electric Fields}
\label{strongE}
The flux of energy carried by an electromagnetic wave is $F=c E^2/4\pi$.  Thus the electric field at the observer associated with a broad-band pulse of flux density $S_\nu$ is \begin{equation}
E_o\sim \left(4\pi S_\nu\nu\over c\right)^{1/2}\sim 10^{-12}\left(S_\nu\over \Jy\right)^{1/2}\left(\nu\over\GHz\right)^{1/2}\esu\, .
\label{EoSnu}
\end{equation}

At separation $r$ from a source at distance $d$, the electric field is larger, $E=(d/r)E_o$.  For $r$ smaller than
\begin{eqnarray}
r_\rel&\sim& \frac{e E_od}{2\pi m_ec\nu}\nonumber\\
&\sim& 10^{13}\left(S_\nu\over \Jy\right)^{1/2}\left(\nu\over\GHz\right)^{-1/2}\left(d\over\Gpc\right)\cm\, ,
\label{rrel}
\end{eqnarray}
the electric field is strong in the sense that
it could accelerate an electron from rest up to relativistic energy on timescale comparable to $(2\pi\nu)^{-1}$.  A free electron would maintain a position of nearly constant phase, in essence surfing on the wave \citep{1969PhRvL..22..728G}.  For $E$ given by Eq.\refnew{EoSnu} and $r\ll r_\rel$, 
the electron would reach a Lorentz factor 
\begin{equation}
\gamma\sim\left(\frac{r_\rel}{r}\right)^{2/3}\, .
\label{gamma}
\end{equation}
Acceleration of electrons in a thermal plasma by a strong broadband pulse would be more complicated.  It is plausible that
the electrons would drag the positive ions along with them to create an outgoing shock wave. Whether this might lead to the emission of coherent  $\GHz$ radio waves is an open question that is best left for a separate investigation. 

It is informative to compare the strength of the electric field near a cosmological FRB with that of giant pulses from the Crab pulsar.  \cite{1999ApJ...517..460S} studied giant pulses in different frequency bands. At $0.6\GHz$, $S_\nu\sim 7000\Jy$  whereas at $1.4\GHz$, $S_\nu\sim 3000\Jy$. Since the Crab is estimated to be at $d\sim 2.2\kpc$ \citep{2005AJ....129.1993M}, the corresponding values of $r_{\rel}$ are a few times $10^9\cm$ in both bands.  These values of $r_\rel$ are about $10$ times larger than the radius of the Crab's light cylinder \citep{2005AJ....129.1993M}, but much smaller than $r_{\rel}$ for FRB 110220.

\section{Discussion \& Conclusions}\label{discussion}
We discuss several properties of FRBs. We conclude that their high DMs cannot be attributed to a stellar corona or a galactic HII region. Thus, if astronomical, they are extra-galactic sources.  We also argue that the propagation through the IGM is unlikely to lead to measurable scatter broadening of $\GHz$ pulses. Thus if scatter broadening is confirmed, it would suggest that the sources are located in dense regions of external galaxies and raise the possibility that a substantial fraction of their DMs are produced there.  

Few sources at $\Gpc$ distances are plausible candidates for producing $\ms$ pulses with $\Jy$ flux densities.  Neutron stars and stellar mass black holes have dynamical timescales of the right order and their gravitational binding energies are more than sufficient.  How the release of binding energy might power a FRB is not clear.
Gravitational waves can be released on $\ms$ timescales, but their coupling to $\GHz$ radio waves is likely to be much slower. Neutrinos carry away most of the binding energy, but only over several seconds \citep{1987PhRvL..58.1494B}.  The sudden release of magnetic energy, perhaps in a giant magnetar flare (\cite{2014arXiv1401.6674L}) or during the collapse of a magnetar into a BH (e.g., \cite{2013arXiv1307.1409F}) seems a better bet.  An advantage of these proposals is that the initial energy is released 
in electromagnetic form. However,  its rapid up-conversion to $\GHz$ 
frequencies poses a hurdle. Whether it can be overcome by the 
acceleration of dense plasma in strong EM fields is questionable.  
Moreover, it is doubtful whether these events occur with sufficient frequency to account for FRBs.

\section*{Acknowledgements}
We thank Sarah Burke-Spolaor, Christian D. Ott and Yanbei Chen for valuable suggestions and comments. We are grateful to an anonymous referee whose advice alerted us to an error in a previous version of section \ref{scattering}.


\end{document}